# Caffeine Modulates the Dynamics of DODAB Membranes: Role of the Physical State of the Bilayer

V. K. Sharma[1,2*], H. Srinivasan[1,2], V. García Sakai[3], and S. Mitra[1,2]

[1]Solid State Physics Division, Bhabha Atomic Research Centre, Mumbai 400085, India

[2]Homi Bhabha National Institute, Anushaktinagar, Mumbai 400094, India

[3]ISIS Pulsed Neutron and Muon Facility, Science and Technology Facilities Council, Rutherford Appleton Laboratory, Didcot OX11 0QX, U.K.

## Abstract

Caffeine (1,3,7-trimethylxanthine), an ingredient of coffee, is used worldwide as a psychostimulant, antioxidant, and adjuvant analgesic. To gain insights into the action mechanism of caffeine, we report on its effects on the phase behaviour and microscopic dynamics of a dioctadecyldimethylammonium bromide (DODAB) lipid membrane, as studied quasielastic neutron scattering (QENS). Tracking the elastic scattering intensity as a function of temperature showed that caffeine does not alter the phase behaviour of the DODAB membrane and that transition temperatures remain almost unchanged. However, QENS measurements revealed caffeine significantly modulates the microscopic dynamics of the lipids in the system, and that the effects depend on the structural arrangement of the lipids in the membrane. In the coagel phase, caffeine acts as a plasticizing agent which enhances the membrane dynamics. However, in the fluid phase the opposite effect is observed; caffeine behaves like a stiffening agent, restricting the lipid dynamics. Further analysis of the QENS data indicates that in the fluid phase, caffeine restricts both lateral and internal motions of the lipids in the membrane. The present study illustrates how caffeine regulates the fluidity of the membrane by modulating the dynamics of constituent lipids depending on the physical state of the bilayer.

*Corresponding Author Email: sharmavk@barc.gov.in; vksphy@gmail.com

# 1. Introduction

Caffeine (1,3,7-trimethylxanthine) is the world's most widely consumed legal psychoactive substance, used to avoid fatigue or to promote attentiveness [1]. It is a small amphiphilic molecule (Fig. 1) that has been widely investigated due to its potential antioxidant activity, exhibiting protective qualities against various human health disorders related to oxidative stress [2]. For example, the antioxidant properties of caffeine have shown protective effects against the development of Alzheimer's disease [3-4]. Caffeine is also used as an additive to enhance pain relief from non steroid anti-inflammatory drugs (NSAIDs), however the exact action mechanism of caffeine is still under debate [1,2,5]. Various mechanisms for caffeine's action, such as mobilization of intracellular calcium, by inhibiting the binding of adenosine and benzodiazepine ligands to neuronal membrane-bound receptors, have been discussed in the literature [1,5]. Caffeine is also found to interact with lipid membranes [5-7]. However, limited information is available on the molecular mechanism of the interaction of caffeine with membrane, which plays a vital role in understanding its actions, accumulation, and metabolism. Furthermore, an understanding of this mechanism will also help in unraveling its impact on the distribution of drugs in the membrane. Recently, a study on unsaturated zwitterionic 1-palmitoyl-2-oleoyl-sn-glycero-3-phosphocholine (POPC) membranes with and without caffeine was carried out using diffraction and molecular dynamic (MD) simulation techniques [5]. It was found that caffeine penetrates into the POPC membranes and mainly locates at the head group–tail group interface of the bilayers. It significantly affects the membrane hydration by attracting water molecules from the membrane itself to form water pockets around itself [5]. The incorporation of caffeine leads to an increase of the membrane thickness, and an overall decrease of the gauche defects in the lipid tails indicating a decrease in the membrane fluidity [5]. In another study using anionic 1-palmitoyl-2-oleoyl-sn-glycero-3-phosphoglycerol (POPG) and zwitterionic 1,2-dioleoyl-sn-glycero-3-phosphocholine (DOPC) bilayers [6-7], it was shown that caffeine partitions in the lipid membrane and accumulates mainly just below the head group region. Furthermore, the effects of caffeine on 1,2-dipalmitoyl-sn-glycero-3-phosphocholine (DPPC)/ dipalmitoyl phosphatidic acid (DPPA) membranes were also studied [8] in the presence and absence of tetracaine, a local anesthetic compound. The study showed that the incorporation of caffeine reduces the fluidization effects of tetracaine. Notwithstanding, the effects of caffeine on the microscopic dynamics of the lipid membranes remain unexplored in detail. Membrane dynamics play a key role in various physiological processes including cell signalling, energy transduction pathways, membrane trafficking, cell division, and bilayer permeability, and therefore, a detailed knowledge of membrane dynamics is necessary for a deep understanding to the action mechanism of caffeine.

The dynamical behaviour of lipids in a membrane is rich encompassing a wide variety of individual molecular motions, like lateral lipid diffusion, lipid flip-flopping, rotational and vibrational motions, as well as collective motions that lead to bending motions and thickness fluctuation of the membrane [9-13]. Such a complex dynamical landscape occurs over several decades in time, ranging from molecular vibrations that occur in the femtosecond scale, to flip-flop of lipids that take place within a few hours. Equally individual molecular motions take place over very short length scales (a few Angstroms), collective motions leading to full membrane mobility span over lengthscales of up to a few microns. A multitude of techniques has been used to probe the dynamics of these systems in such broad temporal and spatial regimes, including nuclear magnetic resonance (NMR) [14], electron paramagnetic resonance (EPR) [15], fluorescence correlation spectroscopy (FCS) [16,17], dynamic light scattering (DLS) [18-19], and quasielastic neutron scattering (QENS) [20-26]. While most of the techniques offer insight into the relaxation timescales of the system, scattering techniques (e.g. QENS and DLS) have the advantage to provide both spatial and temporal information of the relaxation processes in the system. Owing to the wavelength and energy of thermal neutrons, QENS is particularly suited for studying molecular motions in lipid membranes on length scales ranging from Angstroms to nanometers, and on timescales between a few picoseconds and nanoseconds [20-26]. A number of studies have been carried out on self-assembled surfactant and lipid aggregates using QENS, to unravel the dynamics of the constituent molecules in the system [20-30].

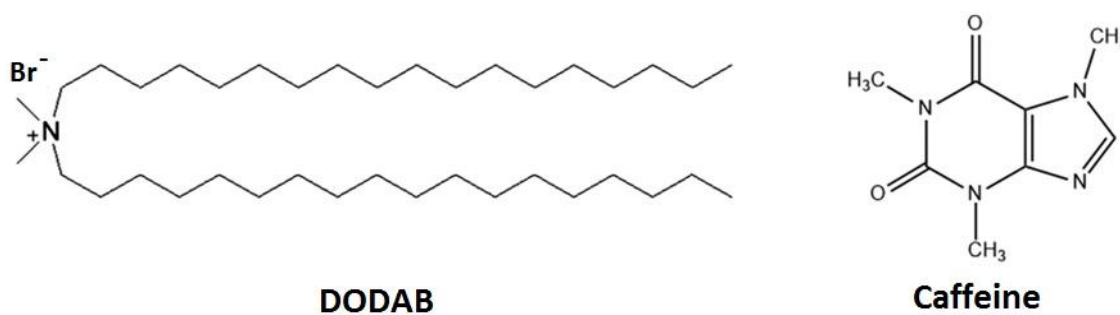

**Fig.1** Chemical structures of the DODAB lipid and caffeine

Dioctadecyldimethylammonium bromide (DODAB) is a double chained cationic lipid (Fig. 1) which forms a lamellar structure in aqueous solution analogous to biological membranes and has been widely investigated due to their membrane mimetic properties [31-36]. Various recent studies have shown the usefulness of DODAB membranes for a number of biomedical applications such as antimicrobials [37-38], drug/vaccine delivery [39-41], and gene/DNA transfection [42]. Besides biomedical applications, DODAB lipids are of fundamental interest due to their rich phase behaviour which depends on temperature, lipid concentration, preparation method and membrane composition [43-45]. We have recently used QENS to investigate the phase behaviour and microscopic dynamics of DODAB lipid membranes [20].

Here, we report the effects of caffeine on the dynamical and phase behavior of a DODAB lipid membrane as studied using QENS. Scanning the temperature dependence of the elastic incoherent scattering signal, through the Elastic Fixed Window Scan (EFWS) method, as well as performing QENS experiments on DODAB lipid membranes with and without caffeine, have revealed various interesting results which indicate a complex interplay between charge, molecular architecture and location of additive within the membrane.

## 2. Materials and Methods
### 2.1 Materials and Sample Preparations

Dioctadecyldimethylammonium bromide [DODAB, $(C_{18}H_{37})_2N(CH_3)_2$ Br] powder (>98%) was purchased from Tokyo chemical industries Co. ltd. Caffeine and $D_2O$ (99.9%) were procured from Sigma Aldrich. 70 mM DODAB vesicles with and without 25 mol % caffeine were prepared by mixing the appropriate amounts of DODAB powder and caffeine in $D_2O$ as described before [20,36]. The mixtures were kept under magnetic stirring for about an hour at ~ 340 K until a clear solution was obtained.

### 2.2 Neutron Scattering Experiments

Two kinds of measurements, namely elastic fixed window scan (EFWS), and quasielastic neutron scattering (QENS), were carried out on DODAB membranes with and without caffeine using the IRIS time-of-flight backscattering spectrometer [46] at the ISIS Pulsed Neutron and Muon Source (Rutherford Appleton Laboratory, UK). The IRIS spectrometer was used in energy offset mode with pyrolytic graphite (002) analyser. This offers an energy resolution of ~ 18 μeV, an energy transfer range of -0.3 to 1.0 meV and a $Q$-range of 0.5 to 1.8 Å$^{-1}$. EFWS experiments were carried out in the temperature range 285 K – 345 K in both heating and cooling cycles. QENS measurements were carried out at 310 K (during heating cycle) and 330K (during cooling cycle), to ensure DODAB lipids were in the coagel and fluid phase, respectively. The temperatures at which QENS data were recorded were chosen based on the phase transitions observed in EFWS. To estimate the solvent contribution, QENS experiments were also carried out on pure D$_2$O at the same temperatures of 310K and 330K. Standard aluminium annular sample holders with an internal spacing of 0.5 mm were used for the neutron scattering measurements, to ensure no more than 10% scattering and minimize multiple scattering effects. A QENS measurement was also carried out on standard vanadium to obtain the instrument resolution. MANTID software [47] was used to perform standard data reduction.

### 3. QENS Data Analysis

QENS measurements provide insights into the correlation of dynamics in the system in the momentum-energy ($Q$, $E$) space. The incoherent scattering component is directly related to self-correlation part and therefore is useful in studying the diffusive motions in the system. The neutron scattering signal from a hydrogenous sample is dominated by the incoherent part due to the tremendously large incoherent scattering cross section of the hydrogen atoms (in comparison to any other atoms). This can be exploited to study the diffusion of lipids in the membrane systems. In aqueous solution, to minimize the scattering contribution from the solvent, D$_2$O is used instead of H$_2$O. Nonetheless, in order to obtain the scattering signal solely from the membrane in the solution the contribution of the solvent is subtracted using the following equation,

$$I_{mem}(Q,E) = I_{solution}(Q,E) - \phi I_{solvent}(Q,E) \tag{1}$$

where, $\phi$ is the fraction of solvent in the system. The subtracted intensity of the lipid membranes is used in the analysis of QENS data.

As already mentioned, lipid membranes exhibit a complex hierarchy of dynamics spanning a wide range of time and length scales. QENS probes dynamics in a temporal regime from nanoseconds to sub picoseconds and at length scales ranging from Angstroms to nanometers. Two distinct motions of the lipids, lateral (~ns) and internal (~ps) motions, fall within these scales and can be observed using QENS [10,20-23]. The experimental data is fitted with the convolution of a model scattering law, $S_{mem}(Q, E)$ and the resolution function of the instrument. $S_{mem}(Q, E)$ is constructed by considering these two dynamically distinct degrees of freedom (lateral and internal) associated with the lipids in the membrane. Assuming these two motions are decoupled, the effective scattering law can be written as a convolution of their individual components,

$$S_{mem}(Q,E) = S_{lat}(Q,E) \otimes S_{int}(Q,E) \tag{2}$$

where, $S_{lat}(Q, E)$ and $S_{int}(Q, E)$ correspond to the scattering laws associated to the lateral and internal motion of the lipids, respectively. Lateral motion pertains to the movement of the whole lipid molecule along the leaflet of the lipid membrane. A variety of models have been suggested for lateral motion of lipids, some of which include, ballistic flow like motion [23], subdiffusion [48], localised diffusion [49], and Fickian diffusion [26]. We have assumed simple Fickian diffusion, based on the recent study [26] which has shown that this is the case for the time and length scales accessible by QENS, and valid at least, for distances greater than a lipid molecule diameter. With the assumption of this model, the explicit scattering law for lateral motion is given as

$$S_{lat}(Q,E) = L_{lat}(\Gamma_{lat}, E) = \frac{1}{\pi} \frac{\Gamma_{lat}}{(E/\hbar)^2 + \Gamma_{lat}^2} \tag{3}$$

where, $\Gamma_{lat}$ is the half-width at half maximum (HWHM) of the Lorentzian corresponding to the lateral motion of lipid. The lateral diffusivity of lipids is obtained by fitting the HWHM with $\Gamma_{lat} = D_{lat}Q^2$. Meanwhile, the general scattering law associated to internal motion of lipids can be written as [10,50],

$$S_{int}(Q,E) = A(Q)\delta(E) + (1 - A(Q))L_{int}(\Gamma_{int}, E) \tag{4}$$

where, $A(Q)$ is the elastic incoherent structure factor (EISF) and $L_{int}(\Gamma_{int}, E)$ is the Lorentzian associated with internal motion. The EISF gives the elastic contribution to the QENS spectra, which arises due to the localised nature of the internal motion of the lipid. The resultant scattering law for the lipid membrane incorporating eqs. (3) and (4) into eq. (2),

$$S_{mem}(Q,E) = \left[ A(Q)L_{lat}(\Gamma_{lat}, E) + (1 - A(Q))L_{tot}(\Gamma_{lat} + \Gamma_{int}, E) \right] \tag{5}$$

### 4. Results and Discussion
### 4.1 Phase Behaviour

In an Elastic Fixed Window Scan (EFWS), the sample temperature is varied in small steps and the elastic intensity (within the energy resolution of the spectrometer) is recorded at each temperature. Increasing the temperature enhances the dynamics in the system, which in turns leads to a loss in the elastic intensity. In case of incoherent neutron scattering, any abrupt change in the elastic intensity is the signature of the phase transition associated with the microscopic dynamics of the system. This technique is very useful to investigate the phase behaviour in lipid membrane systems [20,25]. EFWS measurements were carried out on DODAB membranes with and without caffeine, while heating as well as while cooling, to investigate the effects of caffeine on the phase behaviour. The measured EFWS data is plotted in Fig. 2.

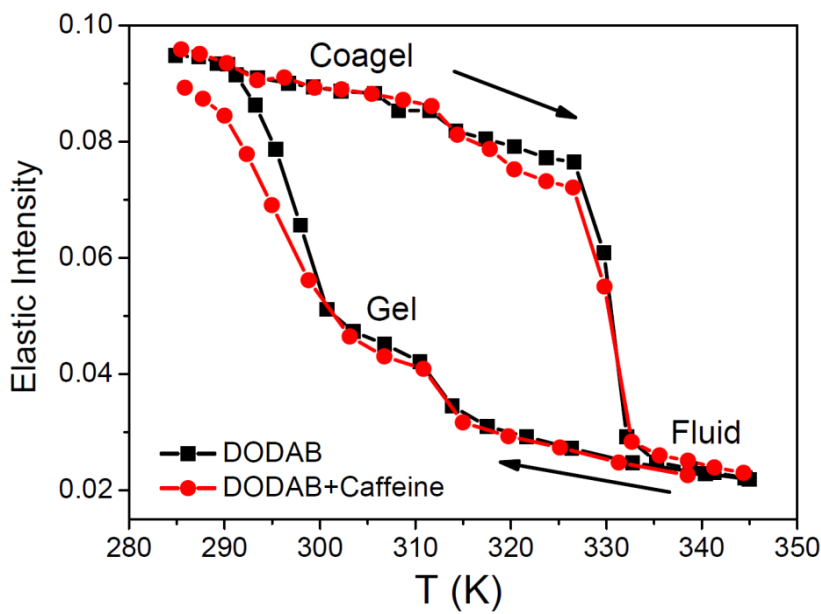

**Fig. 2** $Q$-averaged neutron elastic intensity scans for DODAB vesicles in the absence and presence of caffeine in heating and cooling cycles.

Upon heating, DODAB membranes transition from a coagel to a fluid phase at 327K. However, in the cooling cycle, the fluid phase does not go directly into a coagel phase, but passes through an intermediate gel phase. Moreover, a large hysteresis is observed between the heating and cooling cycles. The existence of an intermediate gel phase in the cooling cycle can be rationalized based on a non-simultaneous change in the dynamics of the polar head groups and the tail groups. The alkyl chains interact mostly via weak van der Waals forces, while a strong electrostatic repulsive interaction exists between the charged head groups. As the temperature is lowered, this causes the hydrophobic alkyl chain tails to order much earlier than the smaller charged head groups, leading to the formation of the gel phase – where the headgroups are still significantly disordered, while the tails are in the ordered state. This behaviour is in strong contrast to that observed in conventional phospholipids such as DMPC, primarily due to differences in the size and polarity of the headgroups [10,28].

The results from the EFWS measurements show that the addition of caffeine does not significantly affect the phase behavior of the DODAB membrane, with the transition temperatures remaining almost unaltered. This is rather different to the effects of non-steroid anti-inflammatory drugs (NSAIDs) on a similar DODAB membrane to that used here (same concentration) [36]. Results from a recent study [36], showed that upon heating, the temperature corresponding to the coagel to fluid phase transition shifts towards a lower temperature, and upon cooling, the NSAIDs inhibit the formation of the intermediate gel phase. The fact that caffeine does not induce similar or any changes to the phase behavior could be due to its exceptionally high $pK_a$ (=14) compared to NSAIDs like aspirin, indomethacin whose $pK_a$ values are in the range 3 to 5 [5,53]. This indicates that at neutral pH, these NSAID molecules are in anionic form in the solution with DODAB, unlike caffeine which mostly in its cationic state. The anionic NSAIDs screen the strong electrostatic repulsion between the positively charged headgroups and effectively cause synchronous ordering between the polar head and hydrophobic tails during cooling, thereby suppressing the formation of the gel phase. On the other hand and due to the high $pK_a$, the caffeine molecule possesses a positive charge and strong affinity to water. Therefore, although it remains near the polar region of the membrane, it does not screen the interactions between headgroups and hence doesn't significantly alter the phase behavior of DODAB membrane. Similar behavior of caffeine has been observed in phospholipid membranes [8]. While the incorporation of NSAIDs lead to a marked effect on their phase behavior, causing a decrease in the main transition temperature [19,25], caffeine does not [8].

### 4.2 Microscopic Dynamics of Lipids

Having established the phase behavior of the DODAB membrane with and without caffeine and that their dynamics in each phase are within the instrument's observation range, we now characterize the molecular motions in detail in these phases. $D_2O$ subtracted QENS spectra for DODAB membrane with and without caffeine at 310K (coagel phase) and 330K (fluid phase) at a representative value of $Q$ of 1.2 Å$^{-1}$ are shown in Figs. 3a and b, respectively. The instrument resolution as measured with a standard vanadium sample, is also shown in Fig. 3a, and to compare the results, the spectra are normalized by their peak amplitude. Significant quasielastic broadening is present for DODAB membrane with and without caffeine at both temperatures measured, suggesting the presence of stochastic motions of the lipids in the system. It is evident from the data that the effect of caffeine on the dynamics of the lipids in the membrane is strongly dependent on the physical state of the membrane. In the coagel phase, the addition of caffeine shows a slight increase in the quasielastic broadening, suggesting an enhancement of the lipid dynamics. On the other hand, its effect is contrary in the fluid phase, as addition of caffeine leads to a decrease in the quasielastic broadening which indicates a stiffening of the membrane.

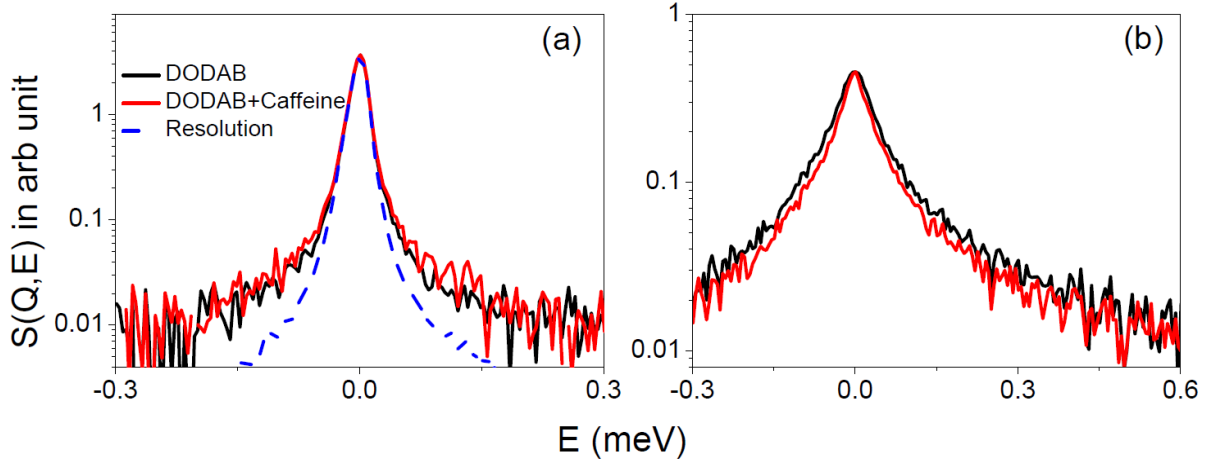

**Fig. 3** Typical observed QENS spectra at a representative $Q = 1.2$ Å$^{-1}$ for DODAB membrane in absence and presence of caffeine in (a) coagel (310K) and (b) fluid (330K) phases. The instrument resolution as measured from a vanadium standard is shown by the dashed line in the left panel. The contribution of the solvent (D$_2$O) has been subtracted, and the resultant spectra are normalized to the peak amplitude for a quantitative comparison.

**4.2.1 Coagel Phase**

The QENS spectra in the coagel phase is analysed by fitting to eq. (4) as shown for a representative $Q = 1.4$ Å$^{-1}$ in Figs. 4a and 4b, for the DODAB membrane without and with caffeine, respectively. The data fits well to the model indicating that only internal motion of the lipids is observable within the resolution of the IRIS spectrometer. This could be due to fact that in coagel phase, lipids are in a densely packed and ordered state, which results in extremely slow lateral motion that is not accessible within the resolution of the spectrometer. The parameters obtained from the fits, the EISF, $A(Q)$, and the HWHM, $\Gamma_{int}(Q)$, are plotted as a function of $Q$ in Figs. 5a and 5b respectively. It is directly evident from the EISF and HWHM in Fig. 5, that the addition of caffeine promotes disorder in the coagel phase of the membrane.

A DODAB molecule consists of two methyl units in the head group and two octadecyl alkyl chains (C$_{18}$H$_{37}$) as the tails, as shown in Fig.1. Thus the internal motions of each lipid molecule can be thought of consisting of the motion of methyl units in the head group and the dynamics of the alkyl tails, separately. Hence, the resulting scattering law for internal motion can be written as a combination of these two components, with their associated weight factors. In this case one can write,

$$S_{int}(Q,E) = P_h S_{head}(Q,E) + P_t S_{tail}(Q,E) \qquad (6)$$

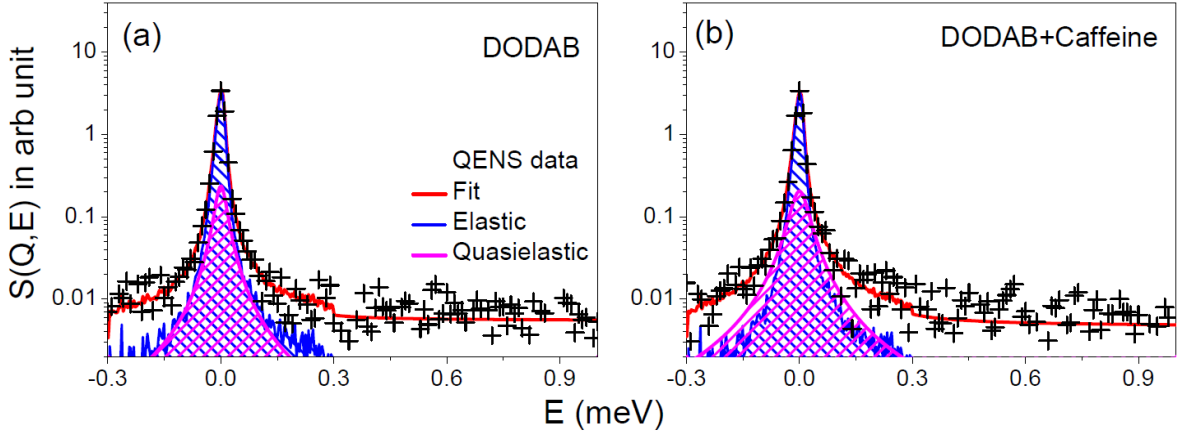

**Fig. 4** QENS spectra in coagel phase (310K) fitted with Eq. (4), at a representative $Q = 1.4$ Å$^{-1}$ for (a) DODAB and (b) DODAB+Caffeine membranes. Individual elastic (blue) and quasielastic (pink) components of the QENS fit are also shown.

where $P_h$ and $P_t$ are the fractions of hydrogen atoms in head group and alkyl tails, respectively. For the DODAB molecule (($C_{18}H_{37}$)$_2$N$^+$(CH$_3$)$_2$Br$^-$), $P_h$ and $P_t$ are equal to 6/80 and 74/80, respectively. Here $S_{head}(Q,E)$ and $S_{tail}(Q,E)$ are the scattering functions corresponding to the headgroup and alkyl tails, respectively. Methyl group dynamics are best described by a *3-fold* jump diffusion model and can be written as [50],

$$S_{head}(Q,E) = \frac{1}{3}[1 + 2j_0(Qa)]\delta(E) + \frac{1}{\pi}\left[\frac{2}{3}[1 - j_0(Qa)]\frac{3\tau_{MG}}{9 + E^2\tau_{MG}^2}\right] \qquad (7)$$

Here, $a$ is the H-H distance (1.8 Å) in the methyl group and $\tau_{MG}$ is the mean residence time of a hydrogen atom in a MG in head.

In coagel phase, lipid molecules are well ordered, and alkyl chains are in the *trans* conformation. In this phase, the alkyl tails could perform uniaxial rotational diffusion along their axis. In the corresponding theoretical model, hydrogen atoms in the alkyl chain undergo reorientations on a circle with a radius of gyration *r*. It has been shown that the scattering law for jump rotations among $N_s$ equivalent sites with large $N_s$ (> 6) and $Qr \leq \pi$, can be used for uniaxial rotational diffusion [51]. Hence, the scattering law for alkyl tails in coagel phase can be given as [51],

$$S_{tail}^{uni}(Q,E) = B_0(Qr)\delta(E) + \frac{1}{\pi}\sum_{n=1}^{N_s-1} B_n(Qr)\frac{\tau_n}{1+E^2\tau_n^2} \quad (8)$$

with

$$B_n(Qr) = \frac{1}{N_s}\sum_{i=1}^{N_s} j_0\left(2Qr\sin\frac{\pi i}{N_s}\right)\cos\frac{2\pi n i}{N_s} \quad (9)$$

and $\tau_n^{-1} = 2\tau^{-1}\sin^2(\frac{n\pi}{N_s})$. Here $j_0$ is spherical Bessel function of the zeroth order and $\tau$ is the average time spent on a site between two successive jumps. In this case, the rotational diffusion constant $D_r$ can be written as:

$$D_r = \frac{2}{\tau}\sin^2\left(\frac{\pi}{N_s}\right) \quad (10)$$

Further, it is possible that at a given temperature, all hydrogen atoms in the alkyl tails might not be mobile within the observation time scale of the spectrometer, and thus only a fraction $p_x$ takes part in the dynamics. Combining all these components together, we write the scattering law for internal motions in the coagel phase as,

$$S_{int}^{coagel}(Q,E) = A_{coagel}(Q)\delta(E) + \frac{1}{\pi}\left[\frac{2P_h}{3}[1-j_0(Qa)]\frac{3\tau_{MG}}{9+\tau_{MG}^2 E^2} + p_x P_t \sum_{n=1}^{N_s-1} B_n(Qr)\frac{\tau_n}{1+\tau_n^2 E^2}\right] \quad (11)$$

where $A_{coagel}(Q)$ is the EISF for the coagel phase and can be written as,

$$A_{coagel}(Q) = \frac{P_h}{3}[1+2j_0(Qa)] + P_t(1-p_x) + P_t\left[\frac{p_x}{N_s}\sum_{i=1}^{N_s} j_0\left(2Qr\sin\frac{\pi i}{N_s}\right)\right] \quad (12)$$

Eq. (12) has been used to describe the observed EISF for DODAB membrane with and without caffeine in the coagel phase. It is evident that the model mentioned above could describe the data well for both the systems and the obtained parameters; the radius of gyration ($r$) and the mobile fraction ($p_x$) are listed in Table I. A notable rise in both the mobile fraction of hydrogen atoms and the radius of gyration is observed indicating incorporation of caffeine enhances the flexibility of the lipid molecules. The uniaxial rotational diffusion constant, $D_r$, and the mean residence of the H-atom's 3-fold reorientation, $\tau_{MG}$, are obtained by fitting the HWHM of internal motion ($\Gamma_{int}$) calculated numerically from eq. (11). The fits shown by continuous and dashed lines for DODAB membrane with and without caffeine, respectively, in Fig. 5b, suggest that the considered model is suitable for the system in coagel phase. The values are given in Table I and indicate a moderate increase in $D_r$ and a decrease in $\tau_{MG}$, supporting the idea of an increased disorder of the membrane in the coagel due to caffeine. The coagel phase of the DODAB membrane is the most ordered phase, with almost all the lipids in all-*trans* conformation. The introduction of caffeine in the membrane can perturb the ordering of lipids in this phase and therefore lead to the observed increase in local dynamics of the system. However, the extent of perturbation is significantly small when compared with the effects of NSAIDs such as aspirin or indomethacin [36]. This suggests that caffeine is weakest plasticizing agent among them for DODAB in the coagel phase.

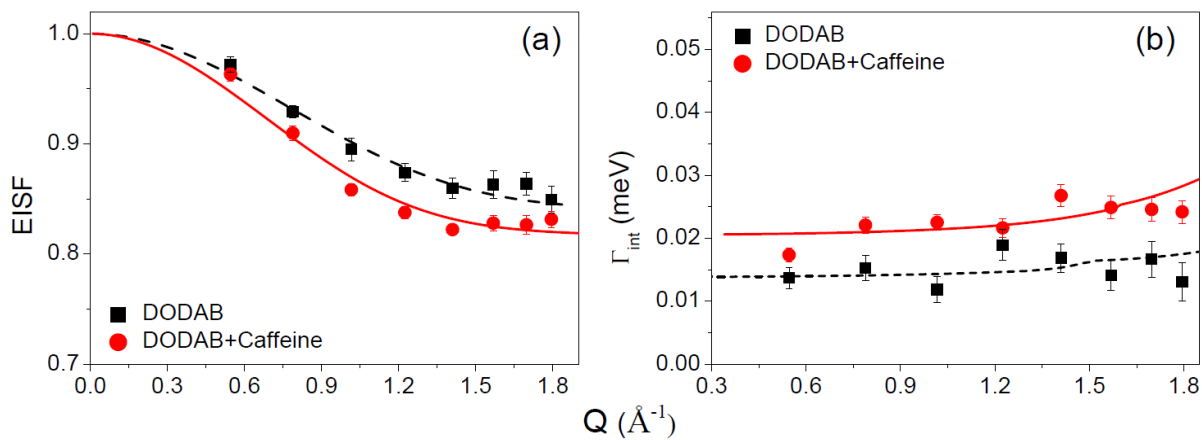

**Fig. 5** (a) EISF and (b) $\Gamma_{int}$ for DODAB in absence and presence of caffeine in the coagel phase (310 K, heating cycles). The solid and dashed lines represent fits assuming fractional uniaxial rotational model as described in the text.

**Table-I** Parameters associated to internal motion of DODAB lipid in the coagel phase (310K in heating cycle) with and without caffeine.

|  | $p_x$(%) | $r$ (Å) | $\tau_{MG}$ (ps) | $D_r$ (×10$^{10}$ s$^{-1}$) |
|---|---|---|---|---|
| DODAB | 13±3 | 1.7±0.2 | 6.7±0.3 | 2.1±0.3 |
| DODAB+Caff | 16±2 | 1.9±0.2 | 6.0±0.3 | 3.0±0.3 |

### 4.2.2 Fluid Phase

The DODAB membrane in the fluid phase exhibits a dynamically richer behaviour with both lateral and internal motion discernible from QENS spectra. The introduction of caffeine in the membrane shows a significant decrease in the lipid dynamics, suggesting that caffeine is a stiffening agent which retards the lipid dynamics in the system in its fluid phase. The QENS spectra of the DODAB membrane with and without caffeine are fitted using eq. (5) accounting for both the lateral and internal motion in the system. The spectra, fits and the individual components (lateral and lateral+internal) are shown at a representative $Q$-value in Figs. 6a and b, for DODAB without and with caffeine, respectively. It is clear from the fits that the timescale of lateral and internal motions of the lipids are well separated.

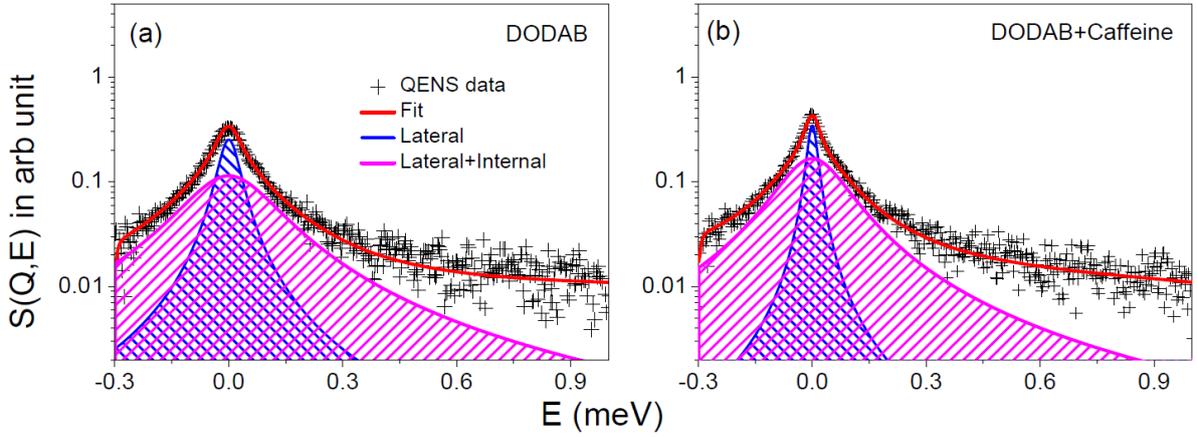

**Fig. 6** QENS spectra in the fluid phase (330K) at a representative $Q = 1.4$ Å$^{-1}$ for (a) DODAB and (b) DODAB+Caffeine membranes, fitted with Eq. (5). Individual components of the QENS fit corresponding to lateral (blue and narrower) and lateral+internal (pink and broader) are also shown.

The HWHM of the lateral motion ($\Gamma_{lat}$) of the lipids for both the cases is plotted as a function of $Q^2$ in Fig. 7, in addition to the linear fits based on Fickian diffusion model ($\Gamma_{lat} = D_{lat}Q^2$). The obtained lateral diffusion constants, $D_{lat}$, for DODAB are listed in Table 2, and evidence the strong effect of caffeine, by a factor of 2. This significant reduction of lateral mobility might be due to local dehydration of lipid membrane interface in the presence of caffeine which strongly hydrates itself at the cost of the membrane [5]. The effect of caffeine on the lateral motion of DODAB lipids in the fluid phase is much more significant than that observed from the common NSAID indomethacin [36]. This indicates that caffeine is the stronger stiffening agent for DODAB in the fluid phase.

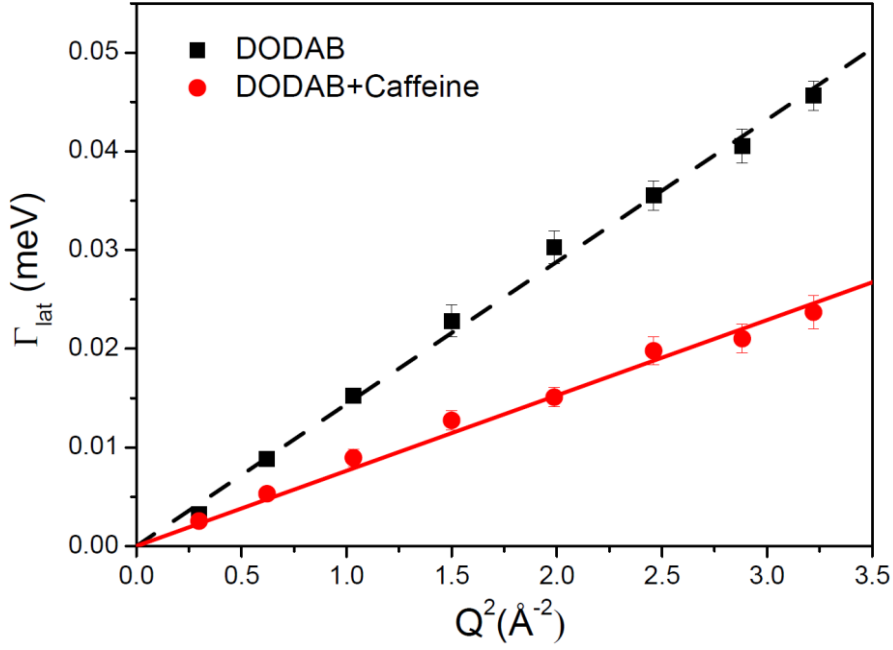

**Fig. 7** Variation of half width at half maximum (HWHM) of the Lorentzian corresponding to lateral motion of lipid, $\Gamma_{lat}$ with $Q^2$ for DODAB membrane with and without caffeine at 330K. The solid and dashed lines represent the Fickian diffusion description as discussed in text.

The fitting parameters $A(Q)$ and $\Gamma_{int}$, used to characterise the internal motion of the lipids in the fluid phase, are obtained from fitting the QENS spectra and are plotted in Figs. 8a and 8b, for DODAB with and without caffeine, respectively. In the fluid phase, the DODAB membrane is noticeably more disordered with significant number of *gauche* defects in the alkyl tails and a large area per lipid molecule. This allows lipids to perform a variety of localised motions such as conformational transitions, bending and stretching motions, chain reorientations, etc. To model this and incorporate the large flexibility of alkyl chains in the fluid phase, a localised translational diffusion (LTD) model with varying radii and diffusivities along the chain is chosen. The scattering law associated to the LTD model is given by [52]

$$S_{tail}^{LTD}(Q,E) = \frac{1}{N_c}\sum_{i=1}^{N_c}\left[A_0^0(QR_i)\delta(E) + \frac{1}{\pi}\sum_{\{l,n\}\neq\{0,0\}}(2l+1)A_n^l(QR_i)\frac{(x_n^l)D_i/R_i^2}{\left[(x_n^l)D_i/R_i^2\right]^2 + E^2}\right] \quad (13)$$

where, $R_i$ and $D_i$ are the radius and diffusivity of $i^{th}$ $CH_2$ units along the chain beginning from the headgroup and $N_C$ is the number of $CH_2$ units in alkyl chain in the lipid – equal to 18 for the DODAB molecule. $A_0^0$ and $A_n^l$ are the elastic and quasielastic structure factors respectively associated to the LTD model [52]. Combining the above equation with the scattering law for the headgroup (eq. 7), we obtain the following expression for $S_{int}(Q, E)$ in the fluid phase,

$$S_{int}^{fluid}(Q,E) = \frac{1}{80}\left[\begin{array}{l}\left\{2(1+2j_0(Qa)) + \frac{74}{N_C}\sum_{i=1}^{N_C}\left[\frac{3j_1(QR_i)}{QR_i}\right]^2\right\}\delta(E) + \\ \frac{1}{\pi}\left\{4(1-j_0(Qa))\frac{3\tau_{MG}}{9+E^2\tau_{MG}^2} + \frac{74}{N_C}\sum_{i=1}^{N_C}\sum_{\{l,n\}\neq\{0,0\}}(2l+1)A_n^l(QR_i)\frac{(x_n^l)^2 D_i/R_i^2}{\left((x_n^l)^2 D_i/R_i^2\right)^2 + E^2}\right\}\end{array}\right] \quad (14)$$

The EISF for the DODAB lipid in the fluid phase can be written as

$$EISF = \frac{1}{80}\left[2[1+2j_0(Qa)] + \frac{74}{N_C}\sum_{i=1}^{N_C}\left[\frac{3j_1(QR_i)}{QR_i}\right]^2\right], \quad (15)$$

but owing to the increasing flexibility of the chain as one goes from the head down along the length of the tail, the radii and diffusivities of $CH_2$ units are considered to increase as it tends towards the tail. A variety of distributions such as linear, log-normal, Gaussian, half-Lorentzian, etc. can be considered for radii and diffusivity [22, 23 45]. In a recent work on molecular dynamics simulations of DODAB bilayer [45], it was observed that distribution of radii and diffusivities in this lipid is half-Lorentzian. In this non-linear distribution, $CH_2$ units near the end of the tail region are significantly mobile compared to the ones near the headgroup. The radius and diffusivity of the $i^{th}$ $CH_2$ unit in this distribution is given by,

$$R_i = \frac{\sigma_R}{\sigma_R^2 + (i-N_C)^2}[R_{max} - R_{min}] + R_{min} \qquad D_i = \frac{\sigma_D}{\sigma_D^2 + (i-N_C)^2}[D_{max} - D_{min}] + D_{min} \quad (16)$$

where $R_{min}$ and $D_{min}$ are the minimum radius and diffusivity which are associated with the $CH_2$ units nearest the headgroup, and $R_{max}$ and $D_{max}$ are the maximum radius and diffusivity associated with the $CH_2$ units at the end of alkyl chain. $\sigma_R$ and $\sigma_D$ are the half-width at half-maxima of the Lorentzians; which characterize the spread of the radii and diffusivity along the alkyl chain.

Localised translational diffusion (LTD) model as described above has been used to describe EISF and HWHM corresponding to internal motions and fits are shown as lines in the Fig.8. It is evident from the figure that LTD model describe the data well. The values of $R_{min}$ and $D_{min}$ are found to very small essentially indicating that the $CH_2$ units near the headgroup show negligible movement. The values of $R_{max}$, $D_{max}$, $\sigma_R$, $\sigma_D$, and $\tau_{MG}$ are given in Table II. Clearly the addition of caffeine leads to a decrease of the diffusivity and an increase in the residence time $\tau_{MG}$, both indicating a stiffening effect on the internal motion of the lipid in the fluid phase. Results are consistent with the recent MD simulation study [5] on POPC membrane which showed that incorporation of caffeine leads to an overall decrease in the gauche defect density indicating decrease in the membrane fluidity. While comparing with indomethacin [36], caffeine has much stronger effect on internal dynamics of DOAB lipids in the fluid phase.

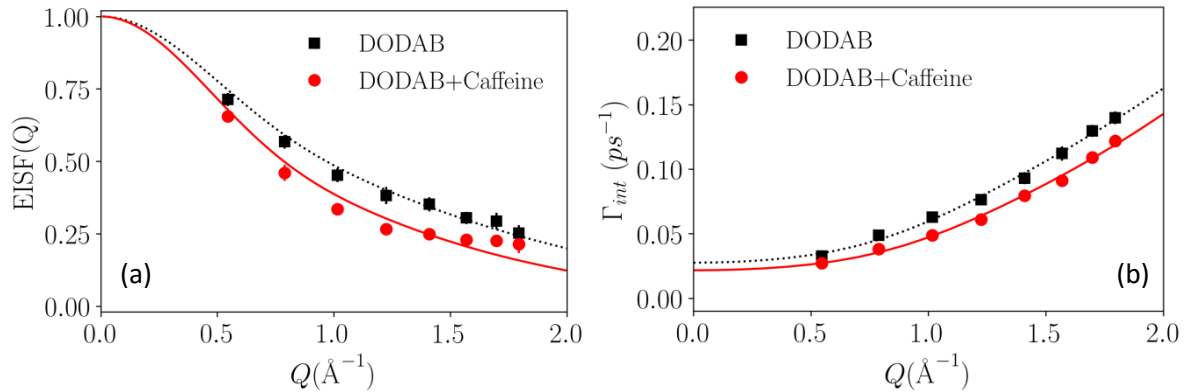

**Fig. 8** Variation with $Q$ of the (a) elastic incoherent structure factor (EISF) and (b) half width at half maximum (HWHM) of the Lorentzian corresponding to the internal motion of lipid, $\Gamma_{int}$ for the DODAB membrane in absence and presence of caffeine at 330K. The solid and dashed lines are the fits according to localised translational diffusion as described in the text.

**Table-II** Parameters associated to lateral and internal motions of DODAB lipid with and without caffeine in the fluid phase (330K).

|  | Lateral motion | Internal motion | | | | |
|---|---|---|---|---|---|---|
|  | $D_{lat}$ (×10$^{-6}$ cm$^2$/s) | $R_{max}$ (Å) | $\sigma_R$ | $D_{max}$ (×10$^{-6}$ cm$^2$/s) | $\sigma_D$ | $\tau_{MG}$ (ps) |
| DODAB | 2.2 ± 0.1 | 4.3 ± 0.2 | 6.0±0.3 | 10.1±0.3 | 2.7±0.2 | 4.5±0.2 |
| DODAB+Caff | 1.2 ± 0.1 | 4.4 ± 0.3 | 8.4±0.1 | 7.9±0.4 | 2.2±0.5 | 5.7±0.4 |

Present study indicates that although caffeine doesn't significantly alter the phase behaviour of the lipid membrane, it affects the dynamics of the lipids in the membrane. The detailed analysis of QENS spectra of DODAB membrane in both coagel and fluid phases reveal that effects of caffeine on the membrane dynamics strongly depend on the physical state of the bilayer. In the coagel, caffeine causes a slight disorder in the membrane, behaving like weak plasticizer. However, in the fluid phase, the effect of caffeine is strong and behaves as a stiffening agent inhibiting both lateral and internal motions of the lipids. Results are compared with the recent studies on caffeine with phospholipid membranes [5,8] and are found to be consistent. Akin to this study, it has been shown that incorporation of caffeine in DPPC/DPPA membrane does not affect the phase behaviour of the membrane [8]. MD simulation study on POPC membrane in fluid phase revealed that addition of caffeine leads to decrease in the membrane fluidity through the formation of local water pockets at the hydrophilic to hydrophobic interface [8]. A similar mechanism could be in play in this present study as well, where the lateral motion of DODAB lipids in fluid phase is strongly inhibited by caffeine. The effects of caffeine on the membrane dynamics in both ordered and fluid phases have been established in this work. The alteration of membrane dynamics can dictate fluidity and permeability, which in turn may considerably modulate the functionality of embedded membrane proteins and transport properties of the cell membrane. This may play an important role in the understanding of action mechanism of caffeine.

## 5. Conclusions

Incoherent elastic intensity scans and quasielastic neutron scattering experiments have revealed interesting effects of caffeine on the microscopic dynamics and phase behaviour of DODAB lipid membranes. The results have been compared to common NSAID's. Elastic fixed window scan measurements showed that the incorporation of caffeine neither shifts the coagel to fluid phase transition in the heating cycle nor inhibits the intermediate gel phase in the cooling cycle. This is in stark contrast to strong effects shown by NSAID's [36]. Further, a detailed analysis of the quasielastic neutron scattering data on DODAB lipid membranes in the absence and presence of caffeine, in coagel and fluid phases, reveals that the addition of caffeine significantly modulates the dynamics of lipid membrane, however with different effects in both phases. In the coagel phase, lipid dynamics are slightly enhanced, whereas in the fluid phase, caffeine restricts lipid mobility quite significantly. Both lateral and internal motions of the lipid are reduced due to the incorporation of caffeine. The results are explained in terms of the charge, molecular architecture, hydration and location of additive within the membrane. The present study sheds some more insight into how caffeine interacts with the membrane and reveals a profound link between the nano-scale structure and the microscopic dynamics of lipids in the membrane. This might be useful to understand the action of other additives of similar chemical and physical properties on membranes.

**References**

1. A. Nehlig, J. L. Daval, and G Debry, *Brain Research Reviews* **17,** 139 (1992).
2. J. R. L. Carmona and A. Galano, *J. Phys. Chem. B* **115**, 4538 (2011).
3. A. Rosso, J. Mossey, C. F. Lippa, *Am. J. Alzheimer's Dis. Other Dementias* **23**, 417 (2008).
4. J. R. P. Prasanthi, B. Dasari, G. Marwarha, T. Larson, X. Chen, J. D. Geiger, O. Ghribi, *Free Radical Biol. Med.* **49**, 1212 (2010).
5. A. Khondker, A. Dhaliwal, R.J. Alsop, Jennifer Tang, M. Backholm, A. C. Shi and M. C. Rheinstädter, *Phys. Chem. Chem. Phys.* **19**, 7101 (2017).
6. M. Paloncýová, K. Berka and M. Otyepka, *J. Phys. Chem. B* **117**, 2403 (2013).
7. M. Paloncýová, R. DeVane, B. Murch, K. Berka and M. Otyepka, *J. Phys. Chem. B*, **118**, 1030 (2014).
8. F.J. Sierra-Valdez, L.S. Forero-Quintero, P.A. Zapata-Morin, M. Costas, A. Chavez Reyes, J.C. Ruiz-Suárez, *PLoS One* **8**, e59364 (2013).
9. J. F. Tocanne, L.D. Ciézanne, A. Lopez, Lateral diffusion of lipids in model and natural membranes, *Prog. Lipid Res.* **33**, 203 (1994).
10. V. K. Sharma, S. Mitra and R. Mukhopadhyay *Langmuir* **35**, 14151 (2019).
11. M. Nagao, E.G. Kelley, R. Ashkar, R. Bradbury, P.D. Butler, *J. Phys. Chem. Lett.* **8,** 4679 (2017).
12. V. K. Sharma, M. Nagao, D.K. Rai, E. Mamontov, *Phys. Chem. Chem. Phys*. **21,** 20211 (2019).
13. D. Marquardt, F. A. Heberle, T. Miti, B. Eicher, E. London, J. Katsaras, et al., *Langmuir* **33**, 3731 (2017).
14. J. Perlo, C. J. Meledandri, E. Anoardo, D. F. Brougham, *J. Phys. Chem. B* **115** (13), 3444 (2011).
15. H. M. McConnell, R. D. Kornberg *Biochemistry* **10**, 1111 (1971).
16. R. Macháň, M. Hof. *Biochimica et Biophysica Acta* **1798**, 1377 (2010).
17. P. Singh, D. Mukherjee, S. Singha, V. K. Sharma, I. I. Althagafi, S. A. Ahmed, R. Mukhopadhyay, R. Das, and S. K. Pal, *RSC Adv.* **9**, 35549 (2019).
18. P. A. Hassan, S. Rana, G. Verma, *Langmuir* **31**, 3 (2015).
19. V. K. Sharma, E. Mamontov and M. Tyagi, *BBA-Biomembrane* **2,** 183100(2020).
20. P. S. Dubey, H. Srinivasan, V. K. Sharma, S. Mitra, V. Garcia Sakai and R. Mukhopadhyay, *Scientific Reports* **8,** 1862 (2018).

21. V. K. Sharma, S. K. Ghosh, P. Mandal, T. Yamada, K. Shibata, S. Mitra and R. Mukhopadhyay, *Soft Matter* **13,** 8969 (2017).

22. V. K. Sharma, E. Mamontov, D. B. Anunciado, H. O'Neill and V. Urban, *J. Phys. Chem. B* **119**, 4460 (2015).

23. S. Busch, C. Smuda, L. C. Pardo, T. Unruh, *J. Am. Chem. Soc.* **132,** 3232 (2010).

24. V. K. Sharma and R. Mukhopadhyay, *Biophysical Reviews* **10,** 721(2018).

25. V. K. Sharma, E. Mamontov, M. Ohl and M. Tyagi, *Phys. Chem. Chem. Phys*. **19**, 2514 (2017).

26. C. L. Armstrong, M. Trapp, J. Peters, T. Seydel, M. C. Rheinstädter, *Soft Matter* **7**, 8358 (2011).

27. J. B. Mitra, V. K. Sharma, A. Mukherjee, V. Garcia Sakai, A. Dash, and M. Kumar, *Langmuir*, **36**, 397 (2020).

28. V. K. Sharma, E. Mamontov, M. Tyagi, S. Qian, D. Rai, V. Urban, *J. Phys. Chem. Lett.*, **7**, 2394 (2016).

29. V. K. Sharma, S. Mitra, G. Verma, P.A. Hassan, V. Garcia Sakai, R. Mukhopadhyay, *J. Phys. Chem. B* **114**, 17049 (2010).

30. V. K. Sharma, D. G. Hayes, V. S. Urban, H. O'Neill, M. Tyagi, E. Mamontov, *Soft Matter*, **13**, 4871 (2017).

31. T. Kunitake, Y. Okahata, *J. Am. Chem. Soc.* **99** (11), 3860 (1977).

32. L. A. Carvalho, A. M. Carmona-Ribeiro, *Langmuir* **14** (21), 6077 (1998).

33. N. Lincopan, E. M. Mamizuka, A. M. Carmona-Ribeiro, *Journal of Antimicrobial Chemotherapy* **52** (3), 412-418 (2003).

34. P. Singh, V. K. Sharma S. Singha, V. G. Sakai, R. Mukhopadhyay, R. Das and S. K. Pal, *Langmuir* **35**, 4682 (2019).

35. A. C. N. Oliveira, T. F. Martens, K. Raemdonck, R. D. Adati, E. Feitosa, C. Botelho, A. C. Gomes, K. Braeckmans, and M. E. C. D. R. Oliveira, *ACS Appl. Mater. Interfaces* **6**, 6977 (2014).

36. P. S. Dubey, H. Srinivasan, V. K. Sharma, S. Mitra V. Garcia Sakai, and R. Mukhopadhyay, *J. Phys. Chem. B* **122**, 9962 (2018).

37. K. P. C. Minbiole, M. C. Jennings, L. E. Ator, J. W. Black, M. C. Grenier, J. E. LaDow, K. L. Caran, K. Seifert, W. M. Wuest, *Tetrahedron* **72** (25), 3559 (2016).

38. A. M. Carmona-Ribeiro, B. V. Debora, L. Nilton, *Anti-Infective Agents in Medicinal Chemistry* **5** (1), 33 (2006).

39. D. B. Vieira, A. M. Carmona-Ribeiro, *Journal of Colloid and Interface Science* **244** (2), 427 (2001).
40. L. Holten-Andersen, T. M. Doherty, K. S. Korsholm, P. Andersen, *Infection and Immunity* **72** (3), 1608 (2004).
41. L. F. Pacheco, A. M. Carmona-Ribeiro, *Journal of Colloid and Interface Science* **258** (1), 146 (2003).
42. J. P. N. Silva, A. C. N. Oliveira, M. P. P. A. Casal, A. C. Gomes, P. J. G. Coutinho, O. P. ; Coutinho, M. E. C. D. Real Oliveira, *Biochim. Biophys. Acta, Biomembr.*, **1808**, 2440 (2011).
43. F.-G. Wu, N.-N. Wang, Z.-W. Yu, *Langmuir* **25** (23), 13394 (2009).
44. F.-G. Wu, Z.-W. Yu, G. Ji, *Langmuir* **27** (6), 2349 (2011).
45. H. Srinivasan, V. K. Sharma, S. Mitra, and R. Mukhopadhyay, *J. Phys. Chem. C* **122**, 20419 (2018).
46. C. J. Carlile, M. A. Adams, *Physica B* **182**, 431 (1992).
47. J. Taylor, O. Arnold, J. Bilheaux, A. Buts, S. Campbell et al. *Bulletin of the American Physical Society* **57**, W26.10 (2012).
48. E. Flenner, J. Das, M. C. Rheinstädter, I. Kosztin, *Phys. Rev. E* **79**, 011907 (2009).
49. U. Wanderlingh, G. D'Angelo, C. Branca, V. C. Nibali, A. Trimarchi, S. Rifici, D. Finocchiaro, C. Crupi, J. Ollivier, H. D. Middendorf, *J. Chem. Phys.* **140**, 174901 (2014).
50. M. Bee, Quasielastic Neutron Scattering; Adam Hilger: Bristol, U.K., (1988).
51. A. J. Dianoux, F. Volino, H. Hervet, *Molecular Physics* **30**, 1181 (1975).
52. F. Volino, A. J. Dianoux, *Molecular Physics* **41**, 271 (1980).
53. M. B. Boggara and R. Krishnamoorti, *Biophys. J.*, **98**, 586 (2010).